\documentclass[twocolumn,showpacs,showkeys,preprintnumbers,amsmath,amssymb,superscriptaddress,pra]{revtex4}
\usepackage[dvips]{graphicx}
\usepackage{dcolumn}
\usepackage{bm}
\usepackage[usenames]{color}
\usepackage{verbatim} 

\newcommand{\affA}{%
  Van der Waals-Zeeman Institute, University of Amsterdam,
  Valckenierstraat 65-67, 1018 XE Amsterdam, The Netherlands}

\newcommand{\affB}{%
  FOM Institute for Atomic and Molecular Physics (AMOLF),
  Science Park 104, 1098 XG Amsterdam, The Netherlands}

\begin{document}

\title{Box traps on an atom chip for one-dimensional quantum gases}

\author{J.~J.~P. van Es}\affiliation{\affA}
\author{P. Wicke}\affiliation{\affA}
\author{A.~H. van Amerongen}\affiliation{\affA}
\author{C. R\'{e}tif}\affiliation{\affB}
\author{S. Whitlock}\affiliation{\affA}
\author{N.~J. van Druten}\affiliation{\affA}

\date{\today}

\begin{abstract} 
We present the implementation of tailored trapping potentials
for ultracold gases on an atom chip. We realize highly elongated
traps with box-like confinement along the long, axial direction combined
with conventional harmonic confinement along the two radial directions.
The design, fabrication and characterization of the atom chip and 
the box traps is described. We load ultracold 
($\lesssim1~\mu$K) clouds of $^{87}$Rb in a box trap, and
 demonstrate Bose-gas focusing as a means to characterize 
these atomic clouds in arbitrarily shaped potentials. 
 Our results show that box-like axial potentials 
on atom chips are very promising for studies of one-dimensional quantum gases.
\end{abstract}



\maketitle

\section{Introduction}
\label{sec:introduction}
 Micro-electromagnetic traps produced using atom chips allow for extremely
tight confinement and precise control of ultracold atoms and quantum degenerate
gases~\cite{FolKruSch02,Rei02,ForZim07}. These chips typically consist of
lithographically defined wire patterns on a substrate which carry modest
currents to create stable and tailor-made magnetic potentials for controlling
atomic motion on the micrometer scale. The success of atom chips is due to the
close proximity of the atoms to the field-producing elements which allow for
high field gradients and tight confinement. For this reason, atom chips are
also versatile tools for fundamental studies of quantum gases in one
dimension~\cite{ReiThy04,PieAigChr07,TreEstWes06,JoChoChr07a,HofLesFis07a,AmeEsWic08,BouDruWes09}. 

The one-dimensional Bose gas exhibits remarkable phenomena that
are not present in either 2D or 3D \cite{PetGanShl04,BloDalZwe08}. Further
interest stems from the availability of
exact solutions for the many-body eigenstates and corresponding
thermodynamics~\cite{LieLin63,YanYan69,KorBogIze93,Tak99,KheGanDru03,YurOlsWei08}.
Almost all experimental work in this direction
 has employed high-aspect-ratio harmonic
potentials. The strong {\em radial} harmonic confinement allows one to treat
the gas as being effectively one-dimensional \cite{Ols98} along the axial
direction, as long as the relevant energy scales are below the radial
excitation energy, $\hbar \omega_\perp$.
However, the {\em axial} harmonic confinement can be disadvantageous, since it leads 
to a spatially inhomogeneous density
distribution, for which a comparison to the above exact theoretical treatments 
necessarily relies on the local-density approximation. 
In contrast, exact solutions do exist for the limiting case of a true box potential with
infinitely steep walls \cite{Gau71,BatGuaOel05}.  The ideal Bose gas also behaves rather
differently in a box when compared to a harmonic trap \cite{GroHooSel50,KetDru96b,DruKet97}.
Furthermore, some interesting features of interacting 1D gases, such
as the quantum decoherent regime  \cite{KheGanDru03,KheGanDru05}, 
are limited to a relatively narrow range of temperatures and densities. 
In an axially harmonic trap, this will typically limit the fraction of the cloud
that is in the regime of interest.
A box-like {\em optical} trap in one dimension was recently 
realized for Bose-Einstein condensates (BECs)~\cite{MeySchHan05}.
Atom chips hold the promise of simpler and more robust access to 
tailored axial potentials by employing specifically 
designed wire patterns~\cite{ReiThy04,PieAigChr07}.

In this paper we demonstrate the experimental implementation of simple box-shaped axial potentials
for studying one-dimensional gases on an atom chip. We describe in some detail
the design, fabrication and characterization of our atom chips. 
These chips have already been successfully used in experiments
on one-dimensional Bose gases 
\cite{AmeEsWic08}, and on the character of radio-frequency-dressed potentials 
\cite{EsWhiFer08}. We extend the technique of 
Bose-gas focusing \cite{ShvBugPet02,RioCoqImp08,AmeEsWic08} to box-like axial
traps, and show that it provides an accurate measure of the energy distribution
of atoms in these traps.

\begin{figure}[t]
\center{\includegraphics[width=0.90\columnwidth]{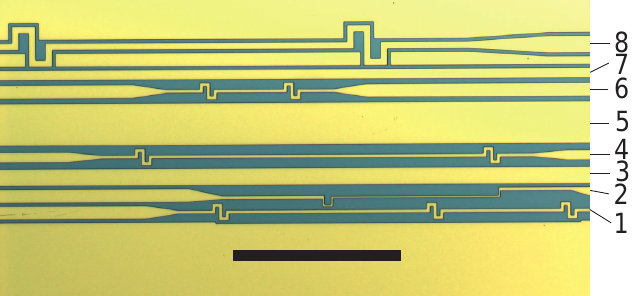}} \caption{(Color
online) Optical microscope image of the central section of an atom chip.
The chip hosts a series of micro-wires of various geometries for creating
one-dimensional traps and box-shaped potentials. The length of the scale
bar is 500~$\mu$m. The numbers at the right identify the atom
chip wires and correspond to the numbers in
Table~\ref{tbl:wires}.}
\label{fig:topview} \end{figure}
The outline of this paper is as follows.
In section~\ref{sec:boxtheory} we discuss the basic trapping geometry
used to produce the box-shaped axial potentials. Typical experimental conditions
allow access to the one-dimensional regime. In
Sec.~\ref{sec:fabrication} we describe in detail the fabrication process
used to realize our atom chips. This section also
includes a description of the various trapping geometries on the chip, including several box traps. In Sec.~\ref{sec:experiments}, we describe experiments with
ultracold atoms loaded into one of the box-shaped traps. This allows for
the resulting magnetic potentials to be characterized.  We analyze
the technique of Bose-gas focusing from box-shaped trapping potentials and 
demonstrate its experimental use.
We discuss some implications of our results and conclude in Sec.~\ref{sec:conclusions}.

\section{Trapping geometry}
\label{sec:boxtheory}
 One of the simplest
trapping geometries for atoms consist of the field of a straight
current-carrying wire combined with a uniform bias field oriented
perpendicular to the wire direction \cite{FolKruSch02,Rei02,ForZim07}. 
The fields cancel at a position
determined by the bias field, producing a quadrupolar field in the
plane orthogonal to the wire, and a minimum field line parallel to
the wire which can be used to guide atoms. The high field gradients
possible in close proximity to the wire (which scale as $r^{-2}$) allow for
tight confinement in the transverse directions.  In order to produce a
trap, full three dimensional confinement is required. This is typically
achieved by bending the wire into a Z-shape, with a long central segment
and two segments at right angles, the latter being referred to as end wires. The end wires
add a spatially varying field component oriented along the central segment
of the wire which
provides longitudinal confinement to the atoms. Typically the longitudinal
potential is weak compared to the transverse confinement resulting in a
highly elongated trap with a trap frequency ratio
($\omega_\perp/\omega_\parallel$) of a few hundred.

The Z-wire geometry has been extremely powerful for preparing ultracold
atoms and Bose-Einstein condensates on atom chips in various experiments
 \cite{FolKruSch02,Rei02,ForZim07}. An important application is in 
generating high-aspect-ratio traps to investigate quantum gases
in the 1D regime 
\cite{TreEstWes06,JoChoChr07a,HofLesFis07a,AmeEsWic08,BouDruWes09}.
The drawback of the Z-wire for this application, however, 
is the inherent harmonic confinement
in the longitudinal direction which limits the possible length of the
studied systems and results in a spatially inhomogeneous atomic density
requiring approximate theoretical treatments. Alternative 
wire geometries have been proposed  to create long box-shaped axial
potentials with nearly constant potential energy at the trap minimum,
combined with tight harmonic confinement in the transverse directions
\cite{ReiThy04,PieAigChr07}.
 The wire patterns we have produced to create box-trap geometries of various dimensions
are shown in Fig.~\ref{fig:topview}. The basic trap is created by a long,
straight and thin wire combined with a perpendicular bias field for
transverse confinement, and two `wiggles' positioned at both ends of the
wire. Each wiggle consists of two opposing notches in the trapping wire
which generate end caps for the trapping potential, eliminating the need
for end wires.

We first describe the field produced by such a box wire geometry. Current
through a single wiggle produces a rapidly decaying magnetic field
component oriented along the wire which produces a good approximation
of a hard-wall potential for
the atoms. The result can be accurately modelled 
by the field of two anti-aligned
magnetic dipoles separated by a distance $\sqrt{2}b$ and with an effective
magnetic moment $Ib^2$ each, where $I$ is the wire current and $b$ is the
characteristic dimension (extension) of the wiggle. For small $b$ the two
dipoles combine to produce a field with quadrupole character:
\begin{equation}
\label{eq:wigglefield} 
\vec{B}_w=
\left(\begin{array}{l} B_x\\ B_y\\ B_z\\\end{array}\right)
 \approx\frac{3\mu_0 I b^3}{4\pi
r^{7}}
\left(
\begin{array}{l} 
z(4x^2-5xy-y^2-z^2)\\ z(x^2+5xy-4y^2+z^2)\\
(y-x)(x^2+y^2-4z^2)\\ 
\end{array}
\right), 
\end{equation}
where $\mu_0=4\pi\times10^{-7}$~Hm$^{-1}$ is the magnetic permeability of
vacuum, $r=\sqrt{x^2+y^2+z^2}$ the distance from the wiggle, $x$ the distance
from the wiggle in the direction of the wire, $y$ is
along the chip surface orthogonal to the wire, and $z$ is the distance
from the chip surface. The potential
experienced by the atoms trapped above the wire ($y=0$) is determined
predominately by the $B_x$ field component which decays as $b^3z/x^5$,
where $b$ and  $z$ are typically $20~\mu$m. This is rapid
compared to the end-wire of a Z-trap (which decays relatively slowly,
as  $z/x^2$),
therefore the wiggle results in a harder potential wall. In a trap the
transverse field components $B_y$, $B_z$ produced by the wiggle are 
much smaller than the applied bias field and simply cause 
small displacements of the quadrupolar field in the $y,z$ plane; these
can be neglected in most applications.

The box-shaped potential is created by positioning two wiggles in a 
long straight wire. In this case the trapping field can be written
as: 
\begin{eqnarray}
\label{eq:boxfield}
\vec{B}_{box}=\vec{B}_w(x-L/2,y,z)+\vec{B}_w(x+L/2,y,z)\notag\\ 
+\vec{B}_l(x,y,z)+\vec{B}_b
\end{eqnarray}
where $L$ is the length of the central segment between the wiggles,
 $\vec{B}_{l}$ is the field of the long wire (excluding the wiggles), 
and $\vec{B}_{b}$ is the bias field. Near
$x=0$, the longitudinal potential consists of a vanishingly small curvature
which scales as $b^3z/L^7$. This should be compared with the longitudinal
potential of a Z-trap with a central segment of length $L$
 which results in a harmonic potential with larger
axial curvature $\propto z/L^4$. The full form of the magnetic field
[Eqs.~\eqref{eq:wigglefield} and \eqref{eq:boxfield}] is relevant for
experiments involving rf-dressed potentials, where the coupling of the rf
field depends on the local orientation of the static field. 
        
Such box traps hold great promise for studying quantum gases in the
one-dimensional regime. To illustrate this we discuss the numbers involved in a
typical experiment. Using $^{87}$Rb atoms in the state $(F,m_F)=(2,2)$, 
(scattering length $a=5.24$~nm)
trapped 20 $\mu$m above a 10-$\mu$m-wide wire carrying 0.5 A of current,
the radial trapping frequency would be $\omega_\perp=2\pi\times30$~kHz, and
the radial size of the ground state $l_\perp=(\hbar/m\omega_\perp)^{1/2}$
would be 62~nm.  The Lieb-Liniger parameter $\gamma=2a/\l_\perp^2 n_1$
\cite{Ols98} that describes the cross-over from the weakly interacting
regime $\gamma<1$ to the strongly interacting regime $\gamma>1$,
would have a value of $1$ at a linear density $n_1=2.7$~$\mu$m$^{-1}$. 
The maximum length of a cloud in the 1D regime (temperature $T$, and 
chemical potential $\mu$ smaller than $\hbar\omega_\perp$) 
can now be estimated by calculating the length over which the axial
trapping potential increases at most  $\hbar\omega_\perp$ with respect
to its minimum.
For wiggles with $b=20$~$\mu$m separated by 1~mm, 
as in wire 4 in Fig.~\ref{fig:topview}, this leads to a length of 
0.73~mm ($\approx 1800$ atoms at $\gamma=1$). 
Further increasing the distance between the wiggles
would allow a proportional increase in the length of the cloud and
in the total number of atoms trapped.
In comparison, for a $Z$-shaped wire with a central segment of 1~mm length, 
the higher harmonic axial confinement leads to an
axial trapping frequency of 17.6~Hz, and  a resulting cloud 
length of $100$~$\mu$m, limiting the number of atoms to below 
180 for a cloud with $n_1=2.7$~$\mu$m$^{-1}$ at its center.

\section{Chip fabrication}
\label{sec:fabrication}
 This section describes
the fabrication of our chip, focusing on the production of long, straight
and high-quality micrometer-sized wires for producing magnetic potentials. The
chips were produced and characterized using the facilities of the Amsterdam
nanoCenter. We describe in some detail the fabrication recipe,
the wire pattern design, characterization of the produced chips, and
the assembly used to support the chip. Further details can be found in
Refs.~\cite{Es09,Ame08b}.

Two key factors played an important role in choosing the
fabrication process. Firstly, we require well-defined and smooth magnetic
potentials which can be accurately calculated from the corresponding
conductor geometry to determine relevant trap parameters. 
The techniques chosen for the fabrication to a
large extent determine the quality of the current-carrying wires and
subsequently the smoothness of the resulting magnetic
potentials~\cite{FolKruSch02,Rei02,ForZim07,EstAusSch04,KruAndWil07}.
 It is also important that the wires are
relatively thin and that the atoms can be trapped close to the surface
where large field gradients provide maximum radial confinement.  Second, we desire
a simple fabrication process in order to save time mastering the procedure
and to allow many chip designs to be implemented in a short time. For this
reason we designed a simple single-layer chip with only a limited number of
wire patterns tailored for experiments with quantum gases in the
one-dimensional regime and with box potentials. The equipment used to
produce these chips is commonly available in any cleanroom facility.

\begin{figure} \center{\includegraphics[width=0.75\columnwidth]{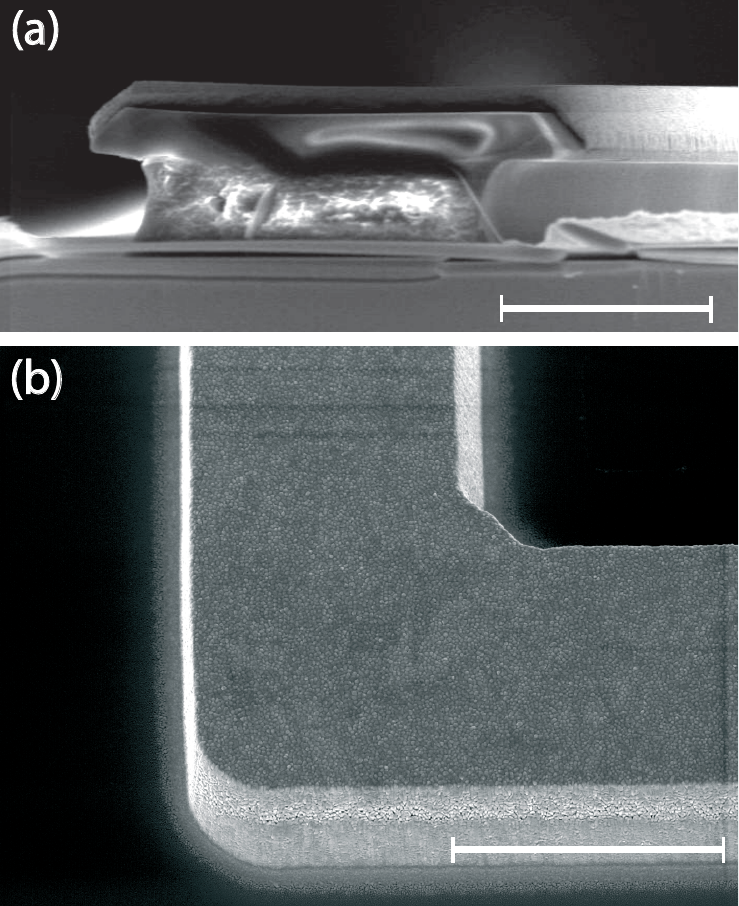}}
\caption{Scanning Electron Microscope (SEM) images of (a) the cross-section
of the resist during fabrication of the atom chip. The undercut is clearly
visible. The image was made immediately after gold evaporation and a
$\sim0.5$-$\mu$m gold layer can be seen lying on top of the resist
structure. (b) a 5-$\mu$m-wide chip wire. The corner displayed is part of a
wiggle used to form a box potential. The scale bars in both figures have a
length of 5~$\mu$m.}
\label{fig:resist}
\end{figure}

We chose a 1.8~$\mu$m-thick evaporated gold layer for the conductors and a
300~$\mu$m-thick silicon substrate. These materials have good thermal
conductivity and the advantage that ample experience processing them is
often available. At the center of the chip we placed a 125-$\mu$m broad
Z-wire with length $L=3~$mm (Fig.~\ref{fig:topview}-wire 5), 
which is used to produce a
deep magnetic trap for initial trapping and evaporative cooling of atoms to
the BEC regime. This wire has a resistance of 0.72~$\Omega$ and is capable
of carrying $\sim2.5$~A continuous current (maximum current density
$\geq10^{6}$ A/cm$^2$). The associated power dissipation of 4.5~W is
sufficiently low so that conduction through the silicon substrate keeps the
temperature below 80~$^\circ$C. As the evaporated gold layer is relatively
thick by microfabrication standards, we use two layers of resist. The top
layer is a photo-sensitive resist patterned using optical lithography. The
bottom layer is a lift-off resist used to create the insulating channels and as a spacer to prevent the top layer from being buried under the gold.
During development the lower resist layer dissolves faster than the top
layer, naturally generating an undercut~(Fig.~\ref{fig:resist}a).
 A layer of gold is
then deposited using metal vapor deposition. By virtue of the overhang the
edges of the deposited gold are well-defined, while the open structure
allows exposure to a solvent. The solvent dissolves the resist, effectively
removing the gold lying on top of it, leaving only the patterned gold layer
on the silicon substrate. Electrical insulation between the chip wires is
through the resistivity of the silicon and the native silicon oxide layer
produced by exposure to air. We chose to keep the chip surface area large
(25$\times$16~mm$^2$ limited by the CF40 vacuum flange) and gold coated to
allow for the reflection of laser beams. The reflective surface is used for
a mirror magneto optical trap (MOT) to collect atoms from background
vapor. The small insulating channels between the wires produce negligible
scattering of the reflected light and do not affect normal MOT operation.

\subsection{Microfabrication recipe}

\emph{1. Substrate preparation.} Substrates are cut from 3-inch silicon wafers
with a (1-1-1) crystal orientation, a thickness of 300~$\mu$m and a
resistivity of $>4000~\Omega$cm. Atom chip substrates were produced from
the wafers by laser cutting to dimensions of $24.63 \times 16.00~$mm$^2$.
We clean the substrate prior to microfabrication in a heated mixture of ammonia and hydrogen-peroxide solution to remove possible Si debris and organic contamination from the surface.

\emph{2. Resist.} The substrate is dehydrated on a hotplate at 150\,$^\circ$C for
5~minutes prior to applying the resist. After the substrate has cooled down
we spin coat (1~minute at 3000~rpm) the bottom resist layer (LOR~20B) to a
thickness of 2.0--2.5~$\mu$m. The substrate is then baked for 5~minutes at
150$^\circ$C. The top resist layer is a positive photoresist
(Microposit~S1813), spin coated (1 minute at 5000~rpm) to a layer of
1.2~$\mu$m thickness. It is then baked for another 30~minutes at 90\,$^\circ$C.

\emph{3. Optical lithography.} The resist is exposed with a dose of $74~$mJ/cm$^2$ of 365-nm near-ultra-violet (NUV) light from a Karl Suss MJB3 mask aligner. A 4-inch Cr mask was made to specification by Deltamask (Enschede, The
Netherlands) using a 442-nm HeCd laser with a resolution of $\sim$1~$\mu$m,
based on an AutoCAD design drawing. We have found that the minimum feature size
of our chips of about 3~$\mu$m is not limited by the lithography mask, but rather by scattering
and divergence of the light inside the thick bilayer resist. After exposure we submerge the substrate in Microposit MF319
developer for 1~minute then rinse in water to stop development. As the
lower layer dissolves faster than the top layer an undercut is produced,
extending $\sim$1~$\mu$m. We then expose the chip surface to a low-energy
oxygen plasma for 5~minutes to remove remaining resist residue from the
exposed silicon.

\emph{4. Gold deposition.} The gold layer is deposited in a PVD (Physical
Vapor Deposition) system at ${\le\!10^{-5}}$~mbar. A 10~nm-thick titanium
bonding layer is first deposited to provide adhesion between the silicon
surface and the gold layer. Gold is then evaporated at a rate of 1~nm/s to
a total thickness of 1.25~$\mu$m, the maximum possible given the crucible
used 
can only hold 3~g of gold at a time. After a refill we continue the
evaporation to 1.8~$\mu$m. The achievable thickness is ultimately limited
by the thickness of the lift-off resist which must prevent filling of the
insulating channels.

\emph{5. Lift off.} We rest the chip in a beaker containing Microposit
Remover~1165 at a temperature of 68$^\circ$C for $\sim$1~hour to lift off
the excess gold. It was helpful to create a bit of flow in the
beaker with a syringe. We do not use ultrasonic agitation as that can
damage the remaining gold wires. After lift off we rinse the chip several
times in acetone and iso-propanol then use the oxygen plasma again to
remove any remaining resist or solvent.

\subsection{Wire pattern design}

The current chip design (Fig~\ref{fig:topview}, Table~\ref{tbl:wires})
 includes 8 wires, all adjacent to
the broad 125-$\mu$m-wide Z-shaped wire which is used for initial trapping
and evaporative cooling. Due to the shape of this wire, all chip wires have
an unavoidable Z shape to prevent crossings in the single conductor layer.
Wires 3 and 7 are regular Z-shaped wires with 3-mm-long straight segments
and widths of 50 and 30~$\mu$m respectively, placed almost symmetrically on
either side of the central Z-wire at a distance of $\sim$100~$\mu$m. These
wires have been used with rf-currents to produce radio-frequency dressed
potentials for manipulating Bose-Einstein condensates~\cite{EsWhiFer08}.
The remaining wires have been designed to produce one-dimensional box
potentials with various lengths defined by the distances between wiggle
segments (Table~\ref{tbl:wires}). Each wire is
separated by 10-$\mu$m-wide insulating channels in the gold layer. At the
center of the chip the exact shape and quality of the wires is crucial
because of the close proximity to the trapped atoms, however outside the
central sections we allow the wires to fan out as much as possible to reduce
wire resistance and ohmic heating. The long edges of the chip host
2.1-mm-wide contact pads which provide ample space to connect each wire to
a contact pin on the chip mount (4.4-mm-wide pads for the 125-$\mu$m Z-wire).
The edges of the chip running parallel to the wires do not have contacts in
order to maximize optical access for imaging.

\begin{table}
\begin{center}
\begin{tabular}{|c|c|c|l|}
\hline
wire & $w$ ($\mu$m) & $R$ ($\Omega$) & function\\
\hline  \hline
1    &   10   & 3.15 & double box structure   \\   
2    &    5   & 3.70 & wiggle test wire   \\
3    &   50   & 1.47 & Z wire, radio-frequency field \\
4    &   10   & 3.89 & 1.00~mm long box \\
5    &  125   & 0.72 & Z wire, initial trapping and cooling \\
6    &   10   & 2.42 & 0.20~mm long box, rf evaporation \\
7    &   30   & 2.06 & Z wire, radio-frequency field \\
8    &   20   & 2.30 & 0.90~mm long box \\
\hline
\end{tabular}
\end{center}
\caption{Characteristics of the wires included on the atom chip shown in
Fig.~\ref{fig:topview}. Widths $w$ and resistances $R$ of the wires are listed, 
along with their intended functions for on-going experiments}
\label{tbl:wires}
\end{table}

\subsection{Chip characterization}
\label{sec:characterization}

\emph{Surface and edge roughness.} 
The quality of the magnetic potentials
produced by microfabricated conducting wires depends critically on both
edge and surface 
roughness~\cite{EstAusSch04,SchEstFig05,ForZim07,PieAigChr07,KruAndWil07}. 
To characterize
these properties we have used a combination of optical microscopy, atomic
force microscopy (AFM), scanning electron microscopy (SEM) and surface
profilometry. We summarize our findings below. Figure~\ref{fig:resist}
 shows images of the resulting wire structures
obtained with a SEM. From this and AFM measurements we find the top surface
and the sides of the wires are smooth down to the grain size of the gold
(typically 50~nm). The observed rounding of the wire corners seen in Figure
2b is the result of the finite resolution of optical lithography. The two
deposition layers of the gold film are also visible as a slight change in
structure along the side of the wire, however, this does not appear to
increase the wire surface roughness.

It is generally difficult to quantify tiny fluctuations of the wire edge
over large length scales which are relevant for experiments with atoms.
Using an optical microscope the wires appear perfectly straight within the
optical resolution of $\sim$1~$\mu$m over length scales of hundreds of
micrometers. We have also measured the flatness of the top surface of a
chip wire over a distance of 800~$\mu$m using a surface profiler (KLA
Tencor Alphastep 500). We find a roughness of 1.8~nm $rms$ with correlations on the 300~$\mu$m length scale. The measured profiles are
similar to those obtained on a bare Si substrate suggesting the evaporated
gold layer accurately follows the substrate surface and has an
approximately uniform thickness. From these measurements we conclude that
the wire patterns are fabricated with high quality and should produce very
smooth potentials for atoms. Ultimately the quality of the wires is
inferred from the magnetic field roughness measured using ultracold atoms
as a probe (Sec.~IVb).

\emph{Chip wire resistance.}
 We have also measured the resistance of the
chip wires under typical experimental conditions. Placed in a vacuum of
$10^{-6}$~mbar to suppress cooling through convection, we performed a
four-terminal measurement of the resistance with a current of about 100~mA,
which is sufficiently low to neglect the increase in resistance due to
ohmic heating. The measured resistances are $\sim10$\% larger than those
expected from a simple calculation, which we attribute to the additional
resistance of the in-vacuum leads and wire bonds. 

Additional measurements of the resistance between different chip wires were
performed to ensure the insulating channels are sufficient to prevent
current leakage under typical operational conditions. Immediately after
fabrication the inter-wire resistances are around 200-600 k$\Omega$,
approximately $10^6$ times larger than the wire resistance. These values
are too large to be explained by the resistivity of the silicon alone,
hence we conclude that the resistance is mainly determined by the thin
native SiO$_2$ surface layer, possibly enhanced by the exposure to the oxygen
plasma during fabrication~\cite{NicBre82}. After three months of use
we found that the inter-wire resistances had decreased significantly to
between 1~k$\Omega$ and 60~k$\Omega$. This is most likely attributable
 to diffusion of
gold into the SiO$_2$ layer. With the new resistances we
anticipate current leakage between wires on the level of $10^{-4}$,
possibly causing deviations from the nominal magnetic fields at the 10~mG
level.  This effect appears to be accelerated by regular heating of the
wires during experiments, as the lowest inter-wires resistances are
measured between those wires used most frequently. After the initial
decrease, the resistances appear to have stabilized at their current values.
We expect this effect may be reduced in the future by incorporating a
dedicated insulating top layer on the Si substrate.

\subsection{Assembly}
 The atom chip is mounted on a macroscopic support
structure which includes six additional wires capable of carrying currents
of up to 20~A for loading atoms to the chip-based microtraps. The
300-$\mu$m-diameter Kapton-coated copper ``miniwires'' run in two isolated
layers oriented along $x$ and $y$ (each with three parallel wires)
positioned at $z=-0.5$~mm and $z=-0.8$~mm underneath
 the chip surface respectively.
The separation between the wires within one layer is 0.65~mm (top layer)
and 3~mm (bottom layer), for the wires along $x$ and $y$ respectively.
These wires fit in grooves in a boron-nitride ceramic disc which has been
machined by a computer-controlled mill. Boron-nitride is easily machined
and has 20 times higher thermal conductivity than Macor.  The miniwires are
electrically connected using standard vacuum-compatible sub-D-type 
gold-plated connector pins. We strip the Kapton from the end of the miniwire and
press it, along with a piece of bare copper, into the male pin. The
resistance of the miniwires (including the connection to the male pins) is
10 m$\Omega$. Operated at a current of 10 A, the power dissipation is 1~W
per wire over its whole length, resulting in negligible heating (10~K/W). 

The Boron-nitride disc is bonded using Epo-tek H77 epoxy to a copper heat
sink mount made of oxygen-free high-conductance (OFHC) copper. This is
bolted to the end cap of a hollow stainless-steel (type 316L) rod with an
outer (inner) diameter of 16~mm (8~mm) which is welded to a CF40 flange for
insertion into the vacuum system. The chip can be water cooled by means of
a polyvinyl chloride (PVC) tube of 6~mm diameter which runs coaxially
inside the stainless-steel rod. We typically run 0.1~l/min of cold tap water through the system during operation. At night this
supply is replaced by heated water at $40~^\circ$C intended to limit the
amount of Rb collected on the chip surface. 

In the final assembly the microchip is carefully glued to the ceramic disc
using Epo-tek 377 epoxy. The alignment error between the on-chip Z-wire and
the miniwires was found to be smaller than $50~\mu$m. The eight chip wires
are connected to the contact pins with $20~\mu$m-diameter aluminium wires
with a wire bonding technique. Each contact pad was bonded with 10 wires
except for the Z-wire where we have used 14 wire bonds to sustain higher
currents. The chip wires are then connected to a set of sub-D-type vacuum
feedthroughs. The maximum bake-out temperature for the chip mount in the
vacuum system is $180^\circ$C, limited by the Kapton coated wires and the
epoxy. The entire chip setup is compatible with ultra high vacuum, and
after bakeout operates at a pressure of $10^{-10}$ mbar.

\section{Ultracold atoms}\label{sec:experiments}

To demonstrate the suitability of the produced potentials for studying ultracold
atoms and quantum gases we have loaded the $L=1~$mm box-shaped trap (wire
4 in Fig.~\ref{fig:topview}).
 This allows the resulting magnetic potentials to be characterized using
the atoms as a probe. We demonstrate evaporative cooling of atoms confined
in the box trap and apply the method of Bose-gas focusing 
\cite{ShvBugPet02,RioCoqImp08,AmeEsWic08}
to perform thermometry of atom clouds.

\subsection{Loading ultracold atoms in the box trap}
A cloud of $^{87}$Rb atoms is initially collected from the background
vapour by a mirror magneto-optical trap positioned below the chip surface.
We then apply a polarization gradient cooling stage and optical pumping to
the atoms in preparation for magnetic trapping. Approximately $2\times10^7$
atoms in the $F=2,m_F=2$ state are caught in a magnetic trap at
$z\approx400$~$\mu$m produced by the central Z-shaped wire ($I_z$=2.25~A) and
an external field ($B_{0y}=13~$G). This field is then ramped up to
$B_{0y}=48$~G over 200~ms to compress the trap and move to $\sim100~\mu$m
from the surface. Next, a logarithmic radio frequency sweep from 27~MHz to
2.1~MHz is applied over 750~ms to evaporatively cool the atoms to a few
microkelvin. The radio-frequency is then increased again to 2.3~MHz prior
to transferring the cloud to the box potential. 

To transfer, we first weaken the confinement of the Z-wire trap and
increase the trap-surface distance to $\sim200~\mu$m by reducing the bias
field to $B_{0y}=20$~G within 50~ms. In 25~ms the Z-wire current is ramped
down to 1~A while the current in the adjacent box-wire is increased to
0.8~A to position the cloud between the two wires. Finally, the Z-wire
current is ramped off completely and the box wire is reduced to 0.37~A with
another 25~ms ramp.

A final cooling stage is applied to the atoms in the box trap using a
two-part rf sweep, first from 2.30~MHz to 1.95~MHz in 60~ms and next
 from 1.95~MHz to
a final frequency between 1.80~MHz and 1.41~MHz in 50~ms. The trap bottom
as measured using rf spectroscopy is at 1.40~MHz. After evaporation the rf
amplitude is switched off completely. Atoms are imaged by turning off the
trap, waiting for a variable time of flight, and then applying a resonant
probe laser pulse to the atoms. The shadow created by the atomic
distribution is then imaged to a CCD camera for analysis. The effective
pixel size in the object plane is $4.3~\mu$m and the optical resolution is
$4~\mu$m.

\begin{figure}[ht]
\center{\includegraphics[width=1\columnwidth]{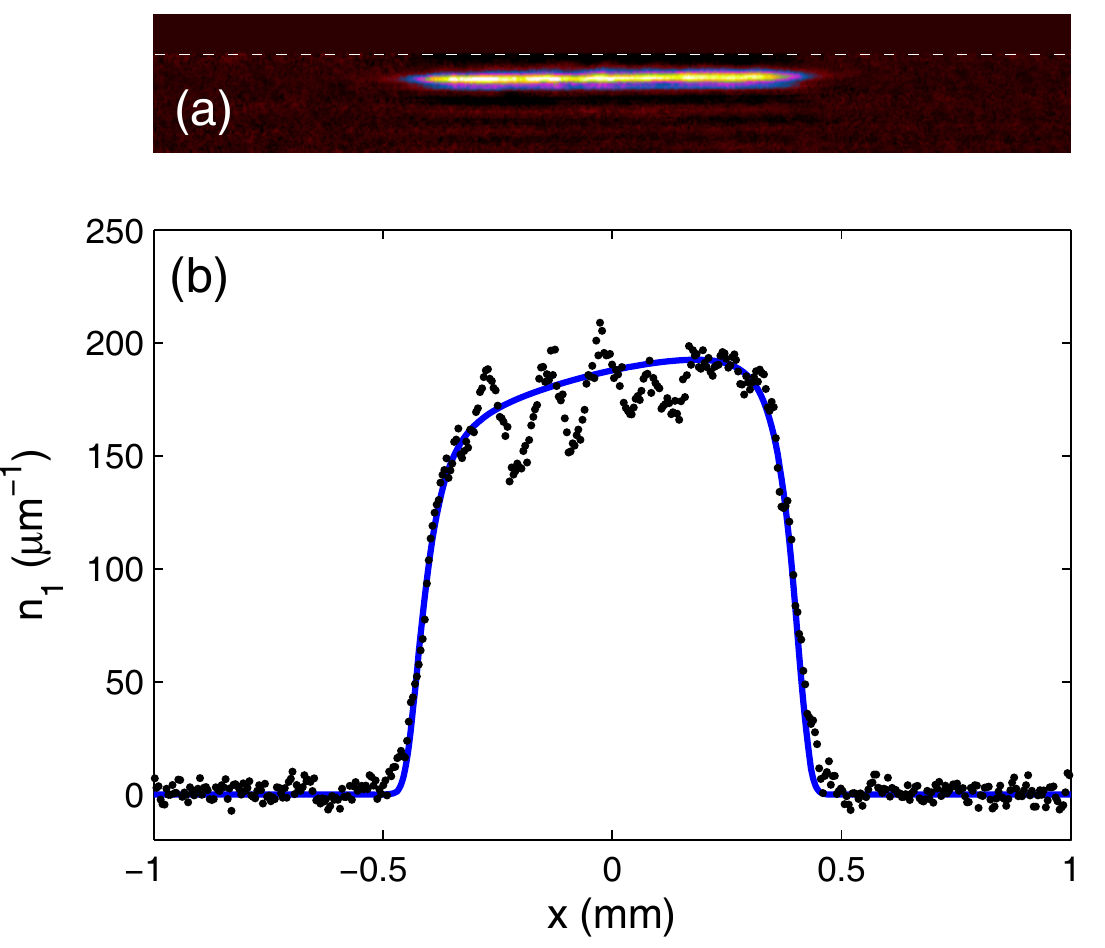}}

\caption{
 (Color online) \emph{In-situ} atomic density distribution after loading the
box-shaped potential for a final evaporation radio frequency of 1.80~MHz.
 (a) Optical density image showing a
$70\times 465$ pixel region ($4.3~\mu$m pixel size in the object plane).
 The dashed line indicates the approximate
position of the chip surface. (b) Integrated profile showing the
corresponding linear atomic density $n_1$ as a function
of longitudinal position $x$. The solid line is a result from the analysis
of Sec.~IVc for the expected in-trap density distribution corresponding to
a temperature of $T=4.16~\mu$K. }
\label{fig:insitu}
\end{figure}

\subsection{Box potential} Figure \ref{fig:insitu}
 shows an optical density image of the
atom cloud after transfer to the box potential for a final rf of 1.80~MHz.
We observe an elongated square-top profile for the atomic distribution as
expected for a Boltzmann gas at thermal equilibrium in  the box-shaped
potential defined by  Eq.~(\ref{eq:boxfield}).
 We note that the distributions are slightly
asymmetric for final radio frequencies above 1.50~MHz (corresponding to
higher cloud temperatures), but become increasingly asymmetric as the radio
frequency is reduced (to lower cloud temperatures). We attribute this to a
small potential gradient, possibly a tilt of the chip surface with respect
to gravity. Additionally, we observe time-independent optical density
modulations in the axial profiles. These however appear roughly independent
of the final evaporation frequency and cloud temperature. This suggests
that the observed modulations are caused partly by artifacts of imaging
close to the chip surface. From the amplitude of the density modulations
and cloud temperatures determined below we infer an upper bound on the
potential roughness of $\Delta V/k_B\leq130$~nK (with $k_B$ Boltzmann's constant)
 or $\Delta B<1.9~$mG \emph{rms}
for a wire current of 0.37~A and a height of $z=40$-$50~\mu$m. The figure of
merit for the box trap is thus $\Delta B/B_{0y}\approx1\times10^{-4}$ which compares well to other chip traps~\cite{EstAusSch04,KruAndWil07}.
 The measured value is consistent with previous measurements on the broader Z-shaped wire
for which we determine $\Delta B/B_{0y}<5\times10^{-5}$ for a height of
$z=120~\mu$m \cite{EsWhiFer08,Es09}.

\subsection{Bose-gas focusing} 
Bose-gas focusing is a powerful experimental
approach for measuring the axial momentum distribution of harmonically
trapped ultracold gases. In particular it has been applied to study
non-equilibrium dynamics of Bose-Einstein condensates \cite{ShvBugPet02} 
and to equilibrium systems in the cross-over from the three-dimensional to the
one-dimensional regimes \cite{AmeEsWic08,Ame08b}.
Here, we extend the description to arbitrarily shaped axial potentials, and apply
this to extract temperatures of atom clouds in box-shaped potentials.  This technique is particularly relevant for one-dimensional gases in box-shaped traps as the spatial distribution of atoms depends weakly on temperature and conventional thermometry methods fail. 

The concept of atom focusing of an elongated cloud is as follows: 
initially, the atoms are in thermal equilibrium in the confining potential. 
A relatively strong parabolic (harmonic) potential is switched on in the axial
direction. This is applied for a short time, typically smaller than
the oscillation period, before being switched off.
 This pulse gives a kick to
the atoms which is proportional to their distance from the trap center and
is analogous to the action of a lens in optics \cite{RioCoqImp08}. Then the atoms are released
from the trap and allowed to evolve during a period of free propagation.
The effect of the kick brings the atoms to an axial focus at a time $t_{\textrm{focus}}$.
At this point the distribution has a minimum axial width proportional to
the cloud temperature and the in-focus distribution reflects the axial
momentum distribution prior to focusing.

In a classical description, suitable for a non-degenerate gas, 
the time evolution of the atomic distribution induced by the focusing
action can be expressed by 
\begin{eqnarray}
g(t)& = & a(t)x_0 + b(t) v_0 \notag\\
h(t) & = & c(t)x_0+d(t)v_0.
\end{eqnarray}
where the final position $g(t)$ and velocity $h(t)$ of an atom depends on
the initial position and velocity $x_0$, $v_0$ through the focus parameters
$a$, $b$, $c$ and $d$.
These parameters  directly correspond to those in the
 $ABCD$ matrix formalism known from conventional (ray- and gaussian-beam) optics
\cite{Sie86}. 
 Transformation of the phase space distribution
$f(x,v,t)$ by this evolution is given by
\begin{equation}
\label{eq:frvt}
f(x,v,t)=\!\int\!\!\!\int\!f(x_0,v_0,0) \delta(x\!-\!g(t))\delta(v\!-\!h(t))dx_0dv_0 
\end{equation}
with $\delta()$ the Dirac delta function. From this it is possible to
determine the real space atomic density distribution $n(x,t)=\nobreak\int
f(x,v,t)dv$. This is conveniently expressed in terms of the initial
distribution
\begin{eqnarray}
n(x,t) &=&\frac{1}{|b(t)|}\int\!f\left(x_0,\frac{x-a(t)x_0}{b(t)},0\right)
    dx_0\label{eq:nrt}
\end{eqnarray}
Thus, we are left with a single integral over the modified initial
phase-space distribution. 

We assume a classical gas in thermal equilibrium given by the 
Boltzmann law:
\begin{equation}
\label{eq:f0}
f(x_0,v_0,0)=C\exp\left[-\frac{1}{k_BT}\left(\frac{1}{2}mv_0^2+V(x_0)\right)\right]
\end{equation}
where $V(x_0)$ is the confining potential energy and $C$ a normalization
constant which yields the total number of atoms. Inserting this initial
distribution in Eq.~\ref{eq:nrt} gives 
\begin{equation}\label{eq:nrtfinal}
n(x,t)=\frac{1}{|b(t)|v_T \sqrt\pi}\int n(x_0) 
    \exp\left[-\left(\frac{x-a(t)x_0}{b(t)v_T}\right)^2\right]dx_0 
\end{equation}
with the thermal velocity $v_T=\sqrt{2k_BT/m}$. The density distribution
during time of flight depends on the initial density distribution
$n(x_0)\!=\!n_0\exp[-(V(x_0)-V_0)/k T]$ where $n_0$ is the peak atomic
density and $V_0$ is the potential energy at the trap minimum.

For a harmonic focusing pulse with a duration $t_p$ and strength $\omega$
followed by a period of free evolution $t$, the relevant evolution
parameters are given by 
\begin{eqnarray}
a(t) &=& \cos(\omega t_p)-\omega t\sin(\omega t_p)\notag\\
b(t) &=& \omega^{-1}\sin(\omega t_p)+t\cos(\omega t_p)
\label{eq:ABCD}
\end{eqnarray}
Equations~\eqref{eq:nrtfinal} and~\eqref{eq:ABCD} describe the
one-dimensional spatial distribution of ultracold atoms during and after
the application of a parabolic focusing pulse for an arbitrary shaped
initial potential $V(x_0)$. In the following, we apply this model to
analyze our experimental results for focusing from the box-shaped potential
of Eq.~\eqref{eq:boxfield}.

\begin{figure}[ht]
\center{\includegraphics[width=1\columnwidth]{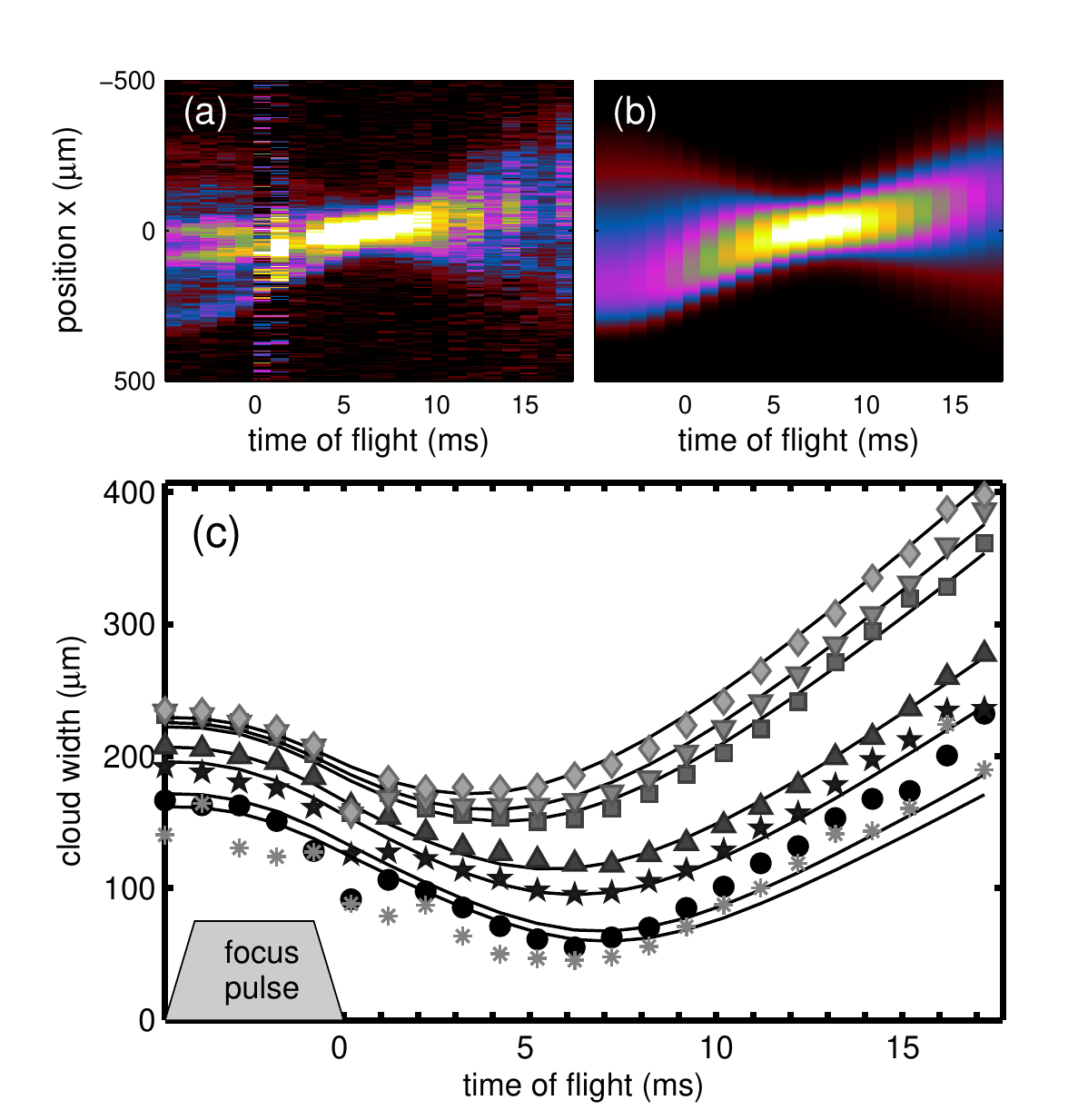}}
\caption{(Color online) Focusing ultracold atoms from a box-shaped potential. (a) Measured
optical density profiles during the focus sequence. (b) Calculated profiles
based on the focusing model of Eq.~\eqref{eq:nrtfinal}. (c) Cloud width
during the focus sequence for several values of the final evaporation radio
frequency and corresponding fits used to determine the cloud temperature.
Symbols (curves) correspond to final radio frequencies (fitted temperatures) of
{\large $\diamond$} 1.80~MHz ($4.16~\mu$K), {$\nabla$} 1.70~MHz
($3.44~\mu$K), {$\Box$} 1.64~MHz ($2.95~\mu$K), {$\triangle$} 1.50~MHz
($1.51~\mu$K), {\large $\star$} 1.46~MHz ($1.02~\mu$K), {\large$\bullet$}
1.42~MHz ($0.42~\mu$K) and {\large$*$} 1.41~MHz ($0.28~\mu$K). The shape
and duration of the focus pulse is indicated by the shaded area.}
\label{fig:focus}
\end{figure}

Results of the focusing experiments are shown in Fig.~\ref{fig:focus}. To
perform the measurements we first increase the 
axial offset magnetic field by 4.25~G within 50~ms. 
To push the atoms away from the chip surface the box-wire current is
increased to 0.95~A in 2~ms. This is
followed by a short trapezoidal current pulse applied to the miniwires
beneath and perpendicular to the box-wire to temporarily introduce a
parabolic axial potential. This pulse consists of a 0.8~ms linear ramp to
0.45~A and 9.8~A for the central and outer-miniwires respectively, held
constant for 4 ms, then finally turned off again in 0.8~ms. The pulse amplitude
corresponds to a peak axial oscillation frequency of $\omega=2\pi\times
25$~Hz. Finally all the fields are switched off ($t=0$), and the atomic
distribution is allowed to freely evolve while falling under gravity. 
During the pulse ($t<0$) or after
a variable time-of-flight ($t>0$) between $t=-4.8$~ms and $t=17.2$~ms
 an optical
density image of the distribution is recorded. The image is then integrated
over the radial coordinate to produce a profile of the linear atomic density.
Profiles for each time-of-flight are then compiled into a
two-dimensional data set $n_1(x,t)$. An example of one such
data set is shown in Fig.~\ref{fig:focus}(a). In total we obtain seven data
sets, each for a different value of the final radio frequency corresponding
to a range of trap depths and final cloud temperatures.

To analyze the data we numerically perform least-squares minimization using
the Nelder-Mead method to the model of Eq.~\eqref{eq:nrtfinal} in order to
determine the temperature $T$ for each value of the final trap depth. The
shape of the initial confining potential is calculated from the known wire geometry. We
use the approximation of Eq.~\eqref{eq:boxfield} with the box length
$L=1040~\mu$m, wiggle size $b=20~\mu$m and trap height $z=50~\mu$m. The
residual harmonic confinement due to the ends of the box-wire corresponding
to a curvature of $1.5~\mu$K/mm$^2$ or $\omega_\parallel=2\pi\times 2.7$~Hz. One
additional parameter specifying an additional gradient to the potential is
required to fully reproduce the observed density distributions and was
determined to be $\Delta=-1.0~\mu$K/mm. This gradient would be consistent
with a tilt of the chip surface with respect to gravity of less than 11~mrad, highlighting the sensitivity of 
such systems to small forces \cite{HalWhiAnd07}.

The focus model provides the expected spatial distribution of atoms before
the focus pulse and agrees well with the measured distribution for each
value of the final radio frequency, see for example Fig.~\ref{fig:focus}(a,b).
 The  focus
pulse is modelled with two parameters, the pulse amplitude [$\omega/2\pi=25$~Hz
in Eq.~(\ref{eq:ABCD})] and a
small displacement between the center of the parabolic potential and the
box potential which causes a small kick to the cloud visible in
 Figs.~\ref{fig:focus}(a,b) as an upward slope in the cloud center of mass.
 Finally fitting is
performed with temperature 
$T$ for each data set as the only free parameter. The
measurements and the result of this fitting procedure are shown in
Fig.~\ref{fig:focus}(a,b) for the 1.50~MHz data set. Good agreement is found for the full range of time-of-flights and the model accurately
reproduces the focus region around $t=6~$ms. The results for all trap
depths are shown in Fig.~\ref{fig:focus}(c), where the cloud width defined
as the second central moment of the distribution is plotted. In this
definition a perfect square-top distribution with length $L=1$~mm equates
to a width of $L/2\sqrt{3}=0.29$~mm. The widths of the observed density profiles 
 are shown with symbols, while the widths of the model distribution are shown with
connected lines. For all data between 1.80~MHz and 1.46~MHz we obtain
excellent agreement between the measured widths and the model distribution.
Some deviation is visible for the lower temperature data sets which may be
attributed to deviations from Boltzmann statistics resulting in a
tighter focus than predicted by the model. For the measurements we
determine the temperatures indicated in the caption of Fig.~\ref{fig:focus}.
 The fitted temperatures are
also consistent with measurements obtained by fitting the
radial expansion rate in order to directly determine the thermal velocity.

\section{Discussion and conclusion}
\label{sec:conclusions}
We have implemented and experimentally investigated simple
box-shaped potentials for trapping ultracold quantum gases on an atom chip. Our
chip hosts a series of long, thin and straight current-carrying wires which
create extremely elongated traps, presently used in studies of Bose gases in
one dimension and in radio-frequency dressed
potentials~\cite{AmeEsWic08,EsWhiFer08}. The described fabrication procedure is
tailored for relatively thick wires for high current capacities and with high
surface and edge quality to produce smooth trapping potentials. We investigate
the quality and suitability of the resulting potentials by loading ultracold
atoms into a 1-mm-long box-shaped trap at a distance of $z\approx 50~\mu$m.
The observed linear density profile has a square-top shape, as expected for a
thermal gas in the box-shaped potential, with small density modulations from
which we infer a potential roughness of $\leq 130$~nK rms. Finally, we have
extended the technique of Bose-gas focusing to arbitrarily shaped axial
potentials and applied the method to measure the energy distribution of thermal
clouds prepared in the box-shaped trap between $0.28~\mu$K and $4.16~\mu$K. We
expect this thermometry technique to be particularly valuable for
one-dimensional systems, as the radial expansion velocity becomes independent
of temperature. Furthermore, in an ideal box trap the spatial distribution of
atoms is independent of the cloud temperature thereby limiting the application
of conventional thermometry techniques. Our results show that box-shaped
potentials implemented on an atom chip are very promising for the study of
one-dimensional quantum gases.

\acknowledgments

We gratefully acknowledge R. J. C. Spreeuw, T. Gregorkiewicz,
G. V. Shlyapnikov and J. T. M. Walraven for helpful discussions. The
atom chip was produced and characterized using the facilities of the Amsterdam
nanoCenter and with the help of J. R\"{o}vekamp. This work is
part of the research program of the Stichting voor Fundamenteel
Onderzoek van de Materie (Foundation for the Fundamental Research on
Matter), and was made possible by financial support from the
Nederlandse Organisatie voor Wetenschappelijk Onderzoek (Netherlands
Organization for the Advancement of Research) and by the European
Union through contract MRTN-CT-2003-505032 (``Atom Chips''), and a Marie Curie fellowship (SW, 
grant number PIIF-GA-2008-220794).

\end{document}